\RequirePackage{lineno}
\documentclass[letterpaper,twocolumn,prl,aps,superscriptaddress,amsmath,amssymb,floatfix]{revtex4-2}
\usepackage{mathptmx}
\usepackage[utf8]{inputenc}
\usepackage{color}
\usepackage{amsmath}
\usepackage{amssymb}
\usepackage{graphicx}
\usepackage{esint}
\usepackage[unicode=true,
 bookmarks=true,bookmarksnumbered=false,bookmarksopen=false,
 breaklinks=false,pdfborder={0 0 1},backref=false,colorlinks=true]
 {hyperref}
\hypersetup{
 linkcolor=magenta,urlcolor=blue,citecolor=blue,pdfstartview={FitH},hyperfootnotes=false}

\makeatletter

\DeclareFontEncoding{LGR}{}{}

\ProvideTextCommand{\~}{LGR}[1]{\char126#1}

\newcommand{\lyxmathsym}[1]{\ifmmode\begingroup\def\b@ld{bold}
  \text{\ifx\math@version\b@ld\bfseries\fi#1}\endgroup\else#1\fi}

\usepackage{textcomp}
\usepackage{epstopdf}

\usepackage{amsfonts}

\pdfpageheight\paperheight
\pdfpagewidth\paperwidth

\@ifundefined{textcolor}{}{%
 \definecolor{BLACK}{gray}{0}
 \definecolor{WHITE}{gray}{1}
 \definecolor{RED}{rgb}{1,0,0}
 \definecolor{GREEN}{rgb}{0,1,0}
 \definecolor{BLUE}{rgb}{0,0,1}
 \definecolor{CYAN}{cmyk}{1,0,0,0}
 \definecolor{MAGENTA}{cmyk}{0,1,0,0}
 \definecolor{YELLOW}{cmyk}{0,0,1,0}
}

\usepackage{xcolor}\usepackage{soul}
\definecolor{blue}{rgb}{0,0,1}
\definecolor{red}{rgb}{1,0,0}
\definecolor{green}{rgb}{0,1,0}

\makeatother

\begin{document}


\title{A cavity QED system with defect-free single-atom array strongly coupled to an optical cavity}
\affiliation{State Key Laboratory of Quantum Optics Technologies and Devices, and
Institute of Opto-Electronics, Shanxi University, Taiyuan 030006,
China}
\affiliation{Collaborative Innovation Center of Extreme Optics, Shanxi University,
Taiyuan 030006, China}

\author{Zhihui Wang}
\thanks{These authors contributed equally to this work.}
\author{Shijun Guan}
\thanks{These authors contributed equally to this work.}
\author{Guansheng Teng}
\author{Pengfei Yang}
\author{Pengfei Zhang}
\author{Gang Li}
\email{gangli@sxu.edu.cn}
\author{Tiancai Zhang}
\affiliation{State Key Laboratory of Quantum Optics Technologies and Devices, and
Institute of Opto-Electronics, Shanxi University, Taiyuan 030006,
China}
\affiliation{Collaborative Innovation Center of Extreme Optics, Shanxi University,
Taiyuan 030006, China}


\begin{abstract}
We experimentally realize a new cavity quantum electrodynamics (QED) platform with defect-free single-atom array strongly coupled to an optical cavity.
The defect-free single-atom array is obtained by rearranging a probabilistically loaded one-dimensional (1D) optical tweezer array with dimensions of $1 \times 40$.
The atom array is enclosed with two cavity mirrors, which compose a miniature optical Fabry-P{\'e}rot cavity with cavity length of 1.15 mm.
By precisely controlling the position of the atom array, we demonstrate uniform and strong coupling of all atoms in the array with the optical cavity.
The average coupling strength between the single atom and the cavity is 2.62 MHz.
The vacuum Rabi splitting spectra for single-atom arrays with atom number $N$ changing from 3 to 26 are measured. 
Thus, the collective enhancement of the coupling strength with ${\sqrt N}$-dependence for multiple atoms is validated at the single atom level.
Our system holds significant potential for establishing the foundation of distributed quantum computing and advancing fundamental research in many-body physics.
\end{abstract}

\maketitle

\section{Introduction}\label{sec1}
The precise control of a single atom in the mode of a microsized optical cavity has enabled the basic research platform, i.e., the cavity quantum electrodynamics (QED) system, to investigate fundamental physics and advance practical applications \cite{RMP2001,RMP2015,RMP2013,RMP20132,RMP2021,Nature2003,science.2004,PRL20022,PRL2019,NP2016,PRL2017,PRX2021,Science2014,Science2020,Nature2021}. 
Recently, the advent of programmable atomic array technologies has demonstrated precise discrimination and control of individual atoms on an unprecedented scale \cite{RMP2010,NP2020,Nature20221,Nature2018,Nature2022,Nature2023,Nature2024}. 
Therefore, a new versatile system can be achieved by integrating the cavity QED system with single-atom arrays. 
The system can significantly improve the flexibility and precision of atom-based cavity QED, particularly in studying many-body physics and the control of quantum processes \cite{PRL2023,PRR2024,PRL20232,PRL20233,QF2024}. 

In such systems, it is possible to realize the arbitrary connections between atoms by proper addressing strategies \cite{PRL20192,Nature20212}. 
This advanced platform opens new opportunities to explore intricate and novel physical phenomena, such as quantum phase transition and collective effect\cite{Nature2020,Nature20242}. 
In addition, it offers a rich resource for harnessing quantum correlations and generating entanglement, which is essential for developing advanced quantum devices. 
Already, it has yielded significant results, including demonstrations of collective enhancement of coupling strength \cite{PRL2023}, the development of multiqubit quantum network registers \cite{Science2024}, the exploration of mid-circuit measurements \cite{PRL2022}, and studies of superradiant and subradiant cavity scattering \cite{PRL20232}. 
These achievements prove the potential of this system to advance both the foundational and applied frontiers of quantum science.

The first key challenge in the development of scalable cavity QED systems is scaling up the size of the coupled atomic array. 
However, the maximum number of atoms in the reported experiments is limited to fewer than 10 \cite{PRL2023,PRL20232,PRR2024,PRL2022,Science2024}, restricting the exploration of many-body effects and the demonstration of large-scale quantum systems. 
The second key challenge is the precise control of the atom-cavity coupling strength for each atom in a large atomic array.
The basic requirement for many theoretical proposals is to realize a uniform coupling for all atoms.
The coupling strength between the atoms and the cavity is sensitive to the relative position between the atom and the standing wave mode of the cavity \cite{PRL2023,PRL2013}. 
As the size of the atomic array increases, the wide spatial distribution of the atomic array increases the complexity to precisely control the position of every atom. 
This resulted spatial variation in coupling strength introduces difficulties in treating all atoms identically for many-body studies and leads to more complex theoretical models.

In this report, we demonstrate an advance in extending the atomic array size and achieving strong coupling between a one-dimensional (1D) defect-free atomic array and a miniature optical cavity. 
Atomic arrays are engineered to couple with the cavity simultaneously with uniform coupling strength. 
We report strong coupling between a cavity and an atomic array with a reconfigurable size, which is rearranged from a 40-tweezer array. 
Up to 26 atoms are manipulated to be strongly coupled with the cavity, and vacuum Rabi splitting spectra of 3 to 26 atoms are observed. 
The ${\sqrt N}$ scaling of the collective enhancement in coupling strength with the deterministic atom number is experimentally validated with more resolved particles, marking a critical step forward in the scalability of cavity QED systems.

\section{Experimental setup}\label{sec2}

\begin{figure}[h]
\centering
\includegraphics[width=\columnwidth]{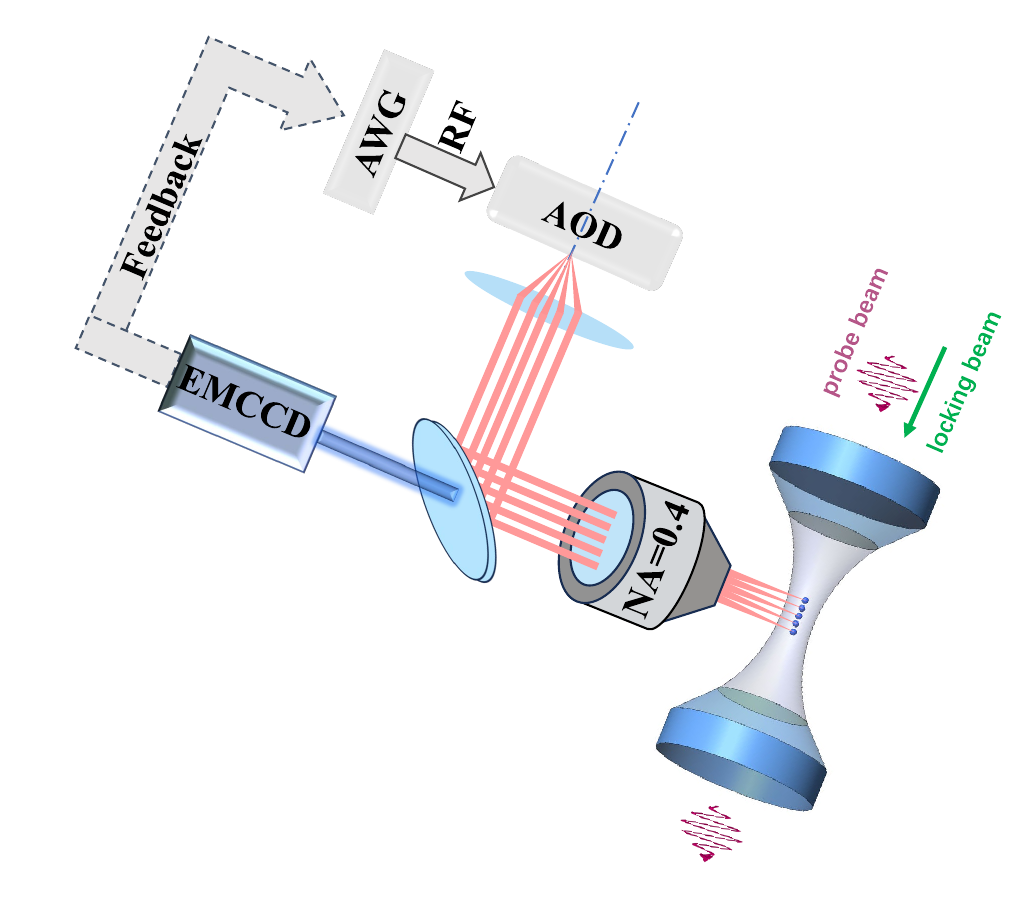}
\caption{ Sketch of the essential part of the experiment setup. Atomic arrays are prepared within an optical cavity and strongly coupled with cavity mode. The length of FP cavity is stabilized by an 851.4 nm locking beam. Another 852 nm probe beam is used to measure the spectrum. }\label{fig1}
\end{figure}

The experimental setup is depicted in Figure \ref{fig1}. 
A high-finesse miniature optical Fabry-P{\'e}rot (FP) cavity composed of two concave mirrors with a curvature radius of 100 mm is employed. 
The cavity has a length of 1.16 mm, and the ${{\rm{TE}}{{\rm{M}}_{00}}}$ mode has a waist size of 45 ${ \mu }$m.
The two mirrors are highly reflective, and the finesse of the FP cavity is ${ 5.8 \times {10^4} }$. 
The cavity length is actively stabilized by an 851.4 nm locking beam and the frequency of one longitudinal mode can be finely tuned around the cesium (Cs) transition line ${ \left|g \right\rangle \equiv \left| {F = 4,m_F = 4} \right\rangle  \leftrightarrow \left| e \right\rangle \equiv \left| {F = 5,m_F = 5} \right\rangle }$. 
Another weak 852-nm laser beam is used to probe the cavity. 
The maximum theoretical coupling strength of a single Cs atom to this cavity is ${ {g_0} = 2\pi  \times 3.4}$ MHz. 
The field decay rates of the cavity and the Cs atom are ${ \gamma  = 2\pi \times 1.1}$ MHz and ${\kappa = 2\pi  \times 2.6}$ MHz, respectively. 
The parameter of cooperativity is then ${ C = {{g_0^2} \mathord{\left/
 {\vphantom {{g_0^2} {(2\kappa \gamma )}}} \right.
 \kern-\nulldelimiterspace} {(2\kappa \gamma )}} = 2}$, which means the cavity achieves the strong coupling regime for a single Cs atom. 

The 1D atom array is obtained by loading and rearranging a 1D optical tweezer array.
A 1064-nm laser beam is firstly transformed to a 1D beam array.
The 1D beam array is generated by driving an acousto-optic deflector (AOD, DTSX, AA Opto Electronic) with a 40-tone radio frequency (RF) signal from an arbitrary waveform generator (AWG, M4i-6631). 
The beam array is then strongly focused by a homemade high numerical aperture (NA) objective with NA = 0.4 and a focal length of f = 28.8 mm \cite{RSI2020} to produce the tweezer array. 
Each tweezer can be independently controlled by the driven RF tone. 
The phase and amplitude of every RF tone are optimized to ensure that the intensity inhomogeneity of all tweezer traps is below 2${ \% }$. 
The trap depth of each tweezer is approximately 0.9 mK with a Gaussian beam waist of ${\approx }$ 1.6 $\mu$m and power of 10 mW. 
The space between adjacent tweezers is set as 4.26 $\mu$m, which is five times the wavelength of the atom transition line. 
The total spatial extent of the 40-tweezer array is 166.1 $\mu$m.
The tweezer array is projected transversely into the cavity from the outside of a vacuum glass cell with the orientation aligned along the cavity axis.

The tweezer array loads single atoms from a cold atom ensemble prepared by a standard magneto-optical trap (MOT) which is aligned in the center of the FP cavity.
The MOT cools and accumulates the cesium atom from an atomic beam emitted from the first-stage two-dimensional MOT.
The atomic ensemble contains approximately ${ {10^5} }$ with a diameter of about 180 µm. 
After a polarization gradient cooling phase, the temperature of the atomic ensemble is approximately 15 µK.
The size of the atomic ensemble is slightly larger than the spatial distribution of tweezer array to ensure that all tweezers load single atom with a uniform probability. 
The tweezers are aligned to overlap with the atomic ensemble and load single atoms by light assistant collision \cite{PRL2002}.
During the loading process, two counterpropagating light beams with polarizations in a Lin ${ \bot }$ Lin configuration are illuminated on the tweezers and induce the inelastic atom collision.
The beams are red detuned from the Cs transition ${ 6{S_{{1 \mathord{\left/
 {\vphantom {1 2}} \right.\kern-\nulldelimiterspace} 2}}}F = 4 \to 6{P_{{3 \mathord{\left/{\vphantom {3 2}} \right. \kern-\nulldelimiterspace} 2}}}F = 5 }$ by ${ 2\pi  \times 24}$ MHz.
At the same time, a repump light resonant with the ${ 6{S_{{1 \mathord{\left/
 {\vphantom {1 2}} \right.\kern-\nulldelimiterspace} 2}}}F = 3 \to 6{P_{{3 \mathord{\left/{\vphantom {3 2}} \right. \kern-\nulldelimiterspace} 2}}}F = 4 }$ transition is applied. 
The fluorescence of the loaded atoms is collected by the same high-NA objective and imaged onto an electron-multiplying CCD (EMCCD) camera. 

The key challenge of the experiment is to precisely control the position of every atom to achieve maximum and stable coupling to the cavity. 
This requires each tweezer to be accurately aligned with both an antinode of the cavity's standing-wave mode along the cavity axis and the center of the transverse mode profile. 
Along the transverse direction, the coupling strength gradually decreases with displacement from the center of the cavity mode, following a Gaussian distribution determined by the mode size. 
Because the cavity mode size (45.3 $\mu$m) significantly exceeds the thermal motion range of the atom ($<1$ $\mu$m) , the variance of coupling strength due to the thermal motion of the atom can be omitted. 
However, along the cavity axis, the coupling strength exhibits a periodic variation due to the standing-wave structure of the cavity mode, with a short period of 426 nm. 
Due to the large size of the optical tweezers (1.6 $\mu$m), the atom cannot be tightly confined around the antinode of the cavity mode with the tweezers alone. 
This limitation can be effectively addressed by employing an additional blue-detuned lattices generated by the 851.4-nm locking beam, which provides an extra confinement along the cavity axis. 
In the blue lattice, the atom trapped in the tweezer will be pushed to a node of the lattice. 
As we designed, the lattice node overlaps perfectly with the antinode of the cavity mode at the right center of the cavity, which guarantees a maximum coupling between the atom and the cavity. 
Due to the slight difference in the wavelengths, the lattice node gradually shifts away from the antinode of the 852-nm cavity mode.
A complete mismatch occurs at a displacement of 193.4 $\mu$m from the cavity center, effectively decoupling the atom from the cavity. 
This overlap and decoupling cycle repeats every 386.8 $\mu$m along the cavity axis. 
However, the atom-cavity coupling strength remains above 90${ \% }$ of that at the cavity center if the atom is trapped within a range of $\pm 56$ $\mu$m from the cavity center.

\section{Experimental results}\label{sec3}

\begin{figure}[h]
\centering
\includegraphics[width=\columnwidth]{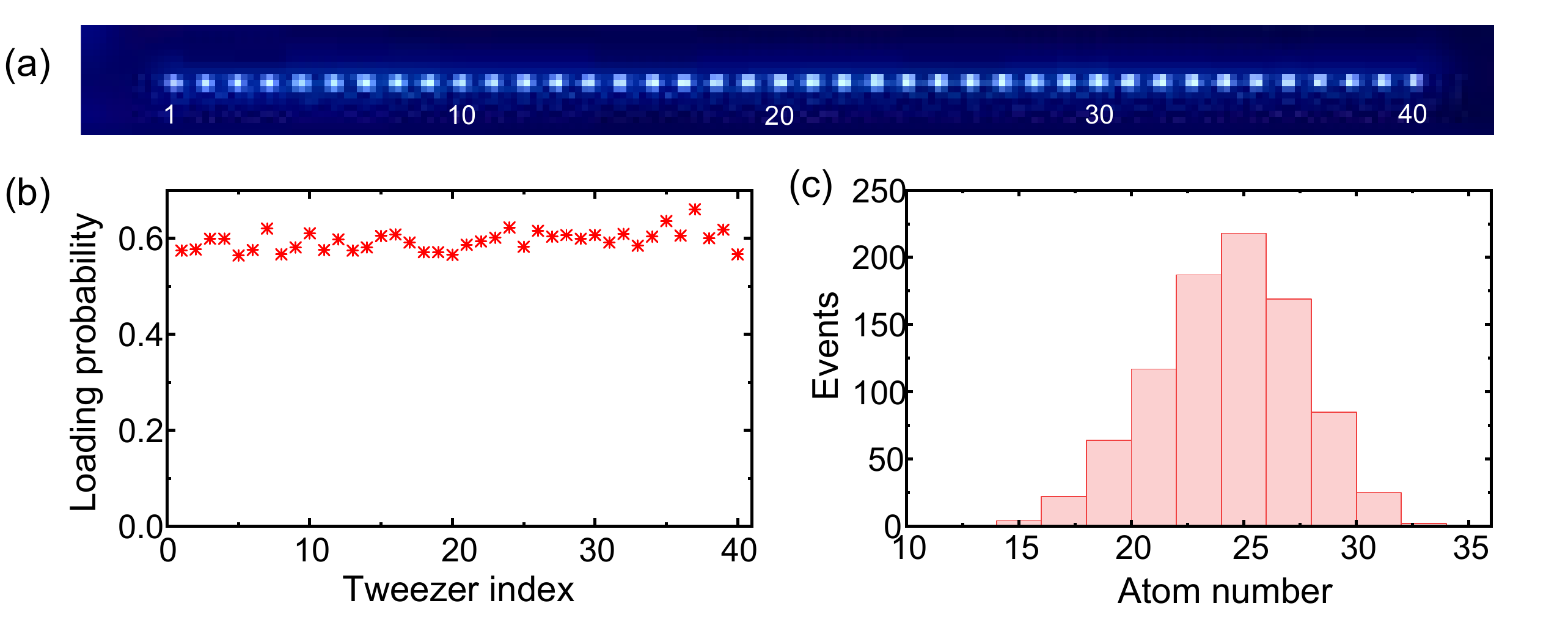}
\caption{ (a) A typical image of single-atom arrays obtained by superimposing approximately 500 loading trials. The exposure time was set as 40 ms for every trial. (b) The initial loading results in an occupation probability of ${\approx }$ 0.6 for each trap in the array. c) Histogram of the number of single atoms loaded into all 40 optical tweezers. The histogram are counted for 890 trials of atom loading, giving an average atom number of approximately 23.8. }\label{fig2}
\end{figure}

In the first step, the loading performance of the tweezer array is tested.
Figure \ref{fig2}(a) shows a typical averaged fluorescence image of single atoms trapped by the tweezer array. 
The distance between neighboring tweezers is set to 4.26 $\mu$m. 
We can see that the single atoms in the trap can be distinctly resolved by our imaging optics.
The distribution of the loading probability for all tweezers is shown in Figure \ref{fig2} (b). 
The loading probability is uniformly distributed and the average value is 0.6, which is  larger than the loading probability of 0.5 by red-detuned collision light in Rubidium experiments.
Thus, an expected mean atom number is approximately 24 for a single loading process with all the 40 tweezers. 
The histogram of the totally loaded atom in all 40 tweezers with 890 experimental trials is shown in Figure \ref{fig2} (c).
We see that the measured mean atom number is 23.8, which confirms the high efficiency of our experiment.

\begin{figure}[h]
\centering
\includegraphics[width=\columnwidth]{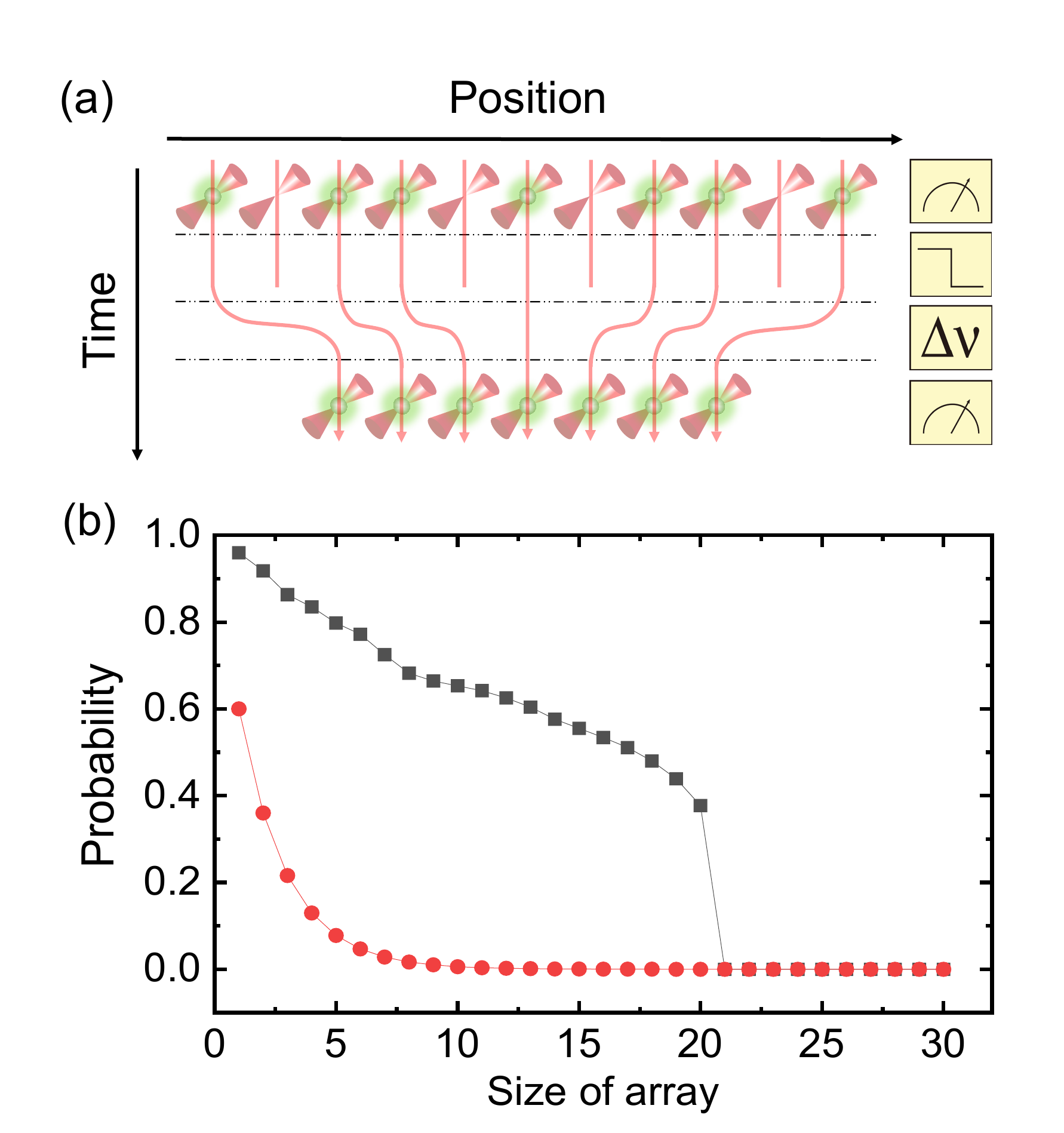}
\caption{ (a) Protocol for creating defect-free arrays. An initial fluorescence image identifies optical tweezers containing single atoms, while empty traps are turned off. The loaded traps are then moved toward the center of the cavity, and a subsequent image is captured to verify the success of the desired atom array. (b) In the initial image, the probability of finding a defect-free length-N array falls off exponentially with N (red dots). After the rearrangement, we demonstrate strongly enhanced success probabilities at producing defect-free arrays (black squares). }\label{fig3}
\end{figure}

Next, an atom rearrangement is used to generate a defect-free atom array \cite{Science2016} and squeeze the overall size of the atom array.
Thus, a better atom-cavity coupling homogeneity for all atoms can be guaranteed. 
The defect-free atom also provides a solid foundation for reliable addressing in future experiments.
Figure \ref{fig3}(a) illustrates the feedback protocol for the rearrangement. 
In the first step, real-time fluorescence imaging yields the information of the trap occupation. 
The unoccupied traps are then turned off by setting the corresponding RF amplitudes to zero. 
In the second step, all the occupied tweezers are moved simultaneously toward the center of the cavity, aligned with the original spacing of 4.26 $\mu$m. 
This rearrangement is achieved by sweeping the RF tones adiabatically within 800 $\mu$s. 
To obtain a defect-free array with the desired atom number, the surplus atom tweezers are switched off in the case that more atoms are loaded or the loading process is repeated in the case that fewer atoms are loaded.
In the third step, a second high-resolution fluorescence image is taken to verify a desired atom array is prepared. 
By rearrangement, the probability of getting a defect-free atom array is greatly enhanced.
The comparison of the results with and without the arrangement is shown in Figure \ref{fig3}(b).
Without rearrangement, the probability of finding a defect-free array of length $N$ is ${{P_N} = {P^N}}$, where ${ P \approx 0.6}$. 
After rearrangement, the probability is dramatically enhanced, especially for the atom array with large atom numbers.
For example, if the desired atom number is 20 in a defect-free atom array, the success probability with the rearrangement is approximately 0.38, which is $10^4$ times the success probability $3.7\times 10^{-5}$ without the arrangement.

\begin{figure}[h]
\centering
\includegraphics[width=\columnwidth]{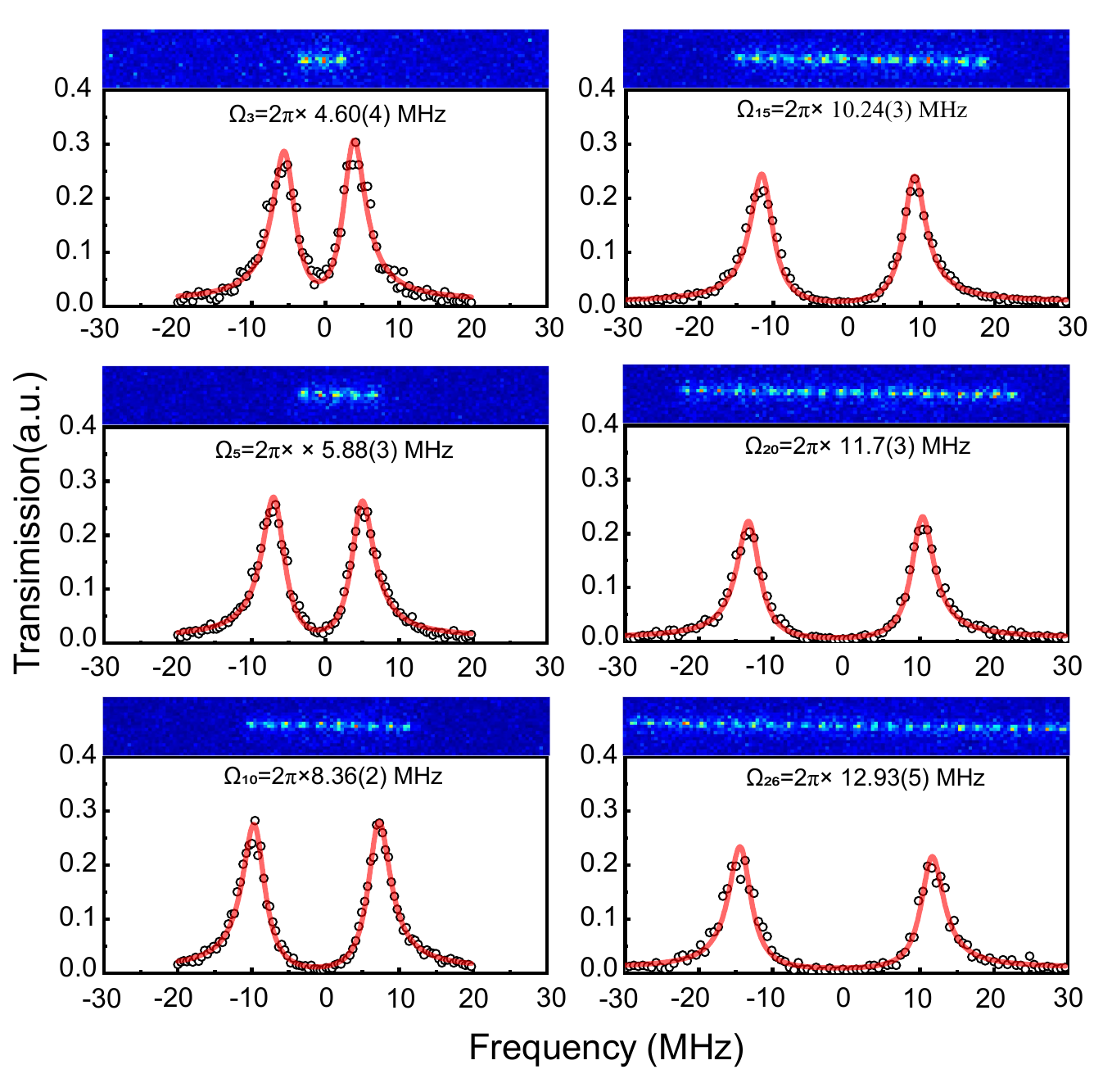}
\caption{ Vacuum Rabi splitting with a deterministic atom number from 3 to 26. The single shot images of the trapped atoms are shown as the inset picture, which are used to precisely count the atom number. The experimental data (black circles) are fitted (red lines) by Eq. (1) to determine ${ {\Omega _N} }$. }\label{fig4}
\end{figure}

After optimizing the position of the defect-free atom array with the 851.4-nm blue lattice, we finally realized uniform coupling between the cavity and atom arrays with atom numbers ranging from 3 to 26. 
The transmissive spectra in the cavity-atom array system and the corresponding single shot images of the atom arrays are shown in Figure \ref{fig4}. 
The atom number is exactly counted from the images.
Clear Rabi splittings can be observed for the defect-free atom arrays, and the coupling strength ${ {\Omega _N} }$ is obtained by fitting the data with the theoretical formula \cite{PRL2023}
\begin{equation}
T = \frac{{{\kappa ^2}\left( {{\gamma ^2} + \Delta _{pa}^2} \right)}}{{{{\left( {{\Omega _{eff}}^2 - \Delta _{pa}^2 + {\Delta _{ca}}{\Delta _{pa}} + \gamma \kappa } \right)}^{^2}} + {{\left( {\kappa {\Delta _{pa}} + \gamma {\Delta _{pa}} - \gamma {\Delta _{ca}}} \right)}^{^2}}}},
\end{equation} \label{eq1}
where ${\Delta _{ca}}$ (${\Delta _{pa}}$ ) is the frequency detuning between the cavity (probe) and atom. 
${ {\Omega _N} }$ is the coupling strength, which is equal to $g$ for a single atom.
${\Delta _{ca}}$ can also be determined by fitting. 
The measured vacuum Rabi splitting of three single atoms is ${ {2\Omega _3}  = 2\pi  \times 9.2}$ MHz, which is approximately 78${ \% }$ of the maximum theoretical value of ${2\pi  \times 11.8}$ MHz. 
The discrepancy is mainly attributed to the imperfections in state initialization. 
Additionally, residual atomic motion contributes to a slight reduction in the coupling strength, resulting in a smaller value. 
The unequal heights of the two normal splitting peaks in Figure \ref{fig4} mainly is because of the variations in ${\Delta _{ca}}$ for different tweezers. 
During the measurements, all tweezers are kept at a shallow trap depth of around 0.1 mK.
Consequently, atomic light shifts fluctuate across tweezers because of small variances in the trap shapes and intensities. 
The ${\Delta _{ca}}$ values extracted from the data fitting range from 0 to 0.4 MHz for all subfigures.

\begin{figure}[h]
\centering
\includegraphics[width=\columnwidth]{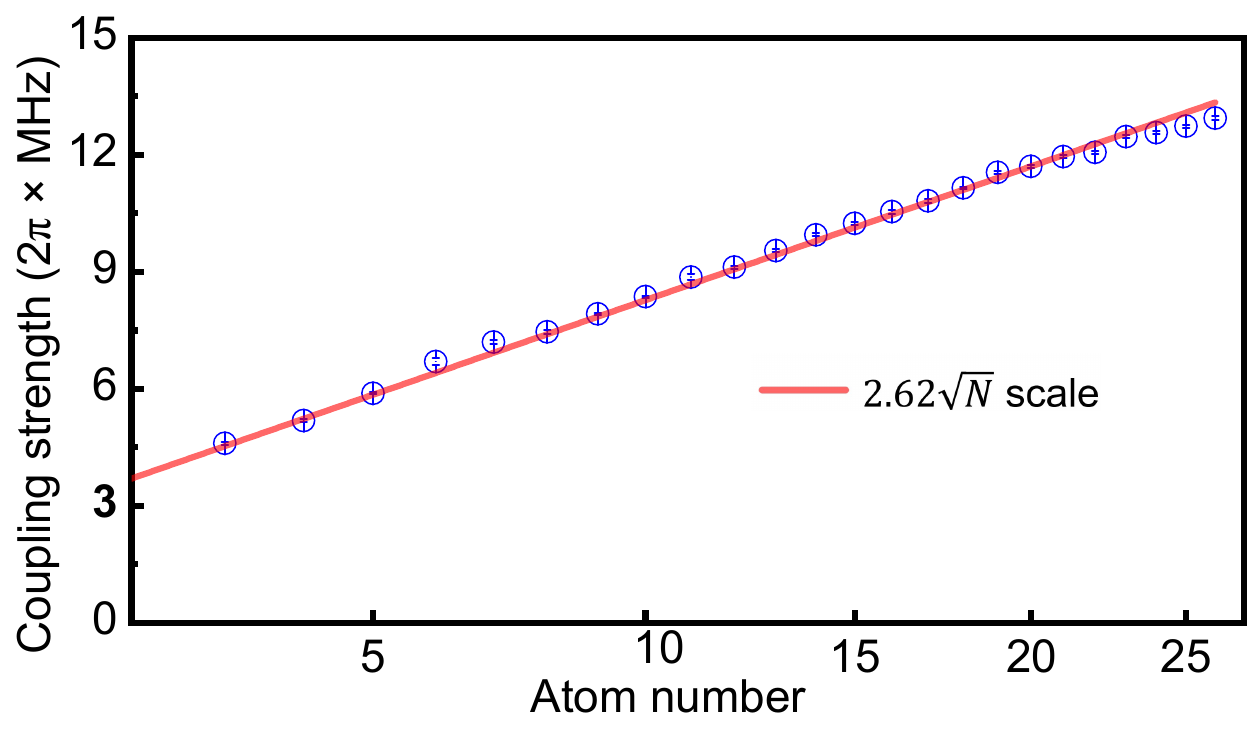}
\caption{ Dependence of the collective coupling strength on the atom number N. The solid red line is the theoretical result for the collective enhancement relation  with the measured single atom coupling strength g=${2\pi  \times 2.62}$ MHz. }\label{fig5}
\end{figure}

In quantum optics, the collective enhancement by using multiple atoms has been accepted as a common principle to increase the light-matter interaction.
The $\sqrt{N}$-dependence of collective enhancement on the resolved atom (qubit) number $N$ has been tested in experiments involving Rydberg excitation of atoms\cite{NP2012,Nature2016} and single qubits in superconducting circuits \cite{PRL2020}.
In our previous study \cite{PRL2023}, the relation was first tested in the optical regime by a cavity QED system with single atom arrays strongly coupled with an optical cavity.
However, due to the small size of the atom array, the maximum atom number is 8.
Here, with the advance in expanding the atom array and the efficient rearrangement, the relation can be tested with 26 atoms.
The experimental result is shown in Figure \ref{fig5}, where the collective coupling strengths are extracted from the data fitting of the vacuum Rabi splittings.
A theoretical relationship for collective enhancement, ${{\Omega _N} = g\sqrt N }$, is also plotted (red line) for comparison with single-atom coupling strength of g=${2\pi  \times 2.62}$ MHz. 
The experimental data agree well with the theoretical prediction with a little description in the large atom numbers.
As the size of the atomic array increases, the collective coupling strength slightly decreases, primarily due to the broader spatial distribution of the atoms even after the rearrangement process. 
Our work provides the most comprehensive demonstration to date with the highest number of distinguishable atoms.

The extracted collective vacuum Rabi splitting ${ {\Omega _N} }$ with $N$ atoms can also be used to assess the homogeneity of the coupling strength for each atoms.
We can deduce that the mean value of the single-atom coupling strength through $g = { \frac{{{\Omega _N}}}{{\sqrt N }}}$. 
Thus, by tuning the atom number from 3 to 26, we get 24 collective Rabi splittings, and then obtain 24 values of $g$. 
The variance in $g$ is found to be within $3.8{ \% }$ of the average value of ${2\pi  \times 2.63}$ MHz. 
Moreover, fitting the data in Figure \ref{fig5} with ${ {g_0} =\frac{{{\Omega _N}}}{{\sqrt N }}}$ yields a single-atom coupling strength of ${ {g_0}  = 2\pi  \times 2.62(1)}$ MHz, which is in good agreement with the averaged value.

\section{Conclusion}\label{sec4}

In summary, we have successfully developed a new cavity QED system in which a well-controlled 1D defect-free atomic array is strongly coupled to a miniature optical cavity.
A defect-free single-atom array with maximum 26 atoms can be efficiently prepared by high-efficient loading of the optical tweezer array with 40 sites and fast rearrangement.
The uniform coupling of each atom to the optical cavity is achieved by carefully aligning the position of the atom in the array with the standing-wave cavity mode by the aid of an intracavity lattice.
Vacuum Rabi splittings for a deterministic number of atoms from 3 to 26 are demonstrated, and the collective enhancement of light-matter interactions is validated in an optical cavity QED system with a maximum atom number of 26.

To increase the size of the atom array, one possible way would be to shrink the distance between adjacent atoms. 
Therefore, the overall size of the array can be enhanced without deteriorating the homogeneity of the coupling strength.
Another way would be to construct a two-dimensional optical tweezer array in the cavity. 
Due to the small variance of the coupling strength along the transverse direction, proper expansion of the atom array in this direction would not deteriorate the homogeneity of the coupling strength either.
Using these methods, the maximum number of atoms in an optical cavity would be increased to hundreds.

Our setup can be extended to study quantum networks in which each node consists of multiple atomic qubits, enabling the development of large-scale quantum information processing systems. 
The strong coupling of individual atoms to the cavity also opens exciting opportunities for exploring many-body physics, where interactions are mediated by photons. 
This work paves the way for further advancements in both foundational quantum science and practical quantum technologies, including quantum computing, quantum communication, and quantum simulation. 

\begin{acknowledgements}
This work was supported by National Key Research and Development Program of China (2021YFA1402002); Innovation Program for Quantum Science and Technology (Grant No. 2023ZD0300400); National Natural Science Foundation of China (U21A6006, U21A20433, 92465201, 12474360 and 92265108); the Fund for Shanxi 1331 Project Key Subjects Construction, and Fundamental Research Program of Shanxi Province (202203021223003).
\end{acknowledgements}

\bibliographystyle{Zou}
\bibliography{decoherence-BlueTrap}

\begin{thebibliography}{40}%
\makeatletter
\providecommand \@ifxundefined [1]{%
 \@ifx{#1\undefined}
}%
\providecommand \@ifnum [1]{%
 \ifnum #1\expandafter \@firstoftwo
 \else \expandafter \@secondoftwo
 \fi
}%
\providecommand \@ifx [1]{%
 \ifx #1\expandafter \@firstoftwo
 \else \expandafter \@secondoftwo
 \fi
}%
\providecommand \natexlab [1]{#1}%
\providecommand \enquote  [1]{``#1''}%
\providecommand \bibnamefont  [1]{#1}%
\providecommand \bibfnamefont [1]{#1}%
\providecommand \citenamefont [1]{#1}%
\providecommand \href@noop [0]{\@secondoftwo}%
\providecommand \href [0]{\begingroup \@sanitize@url \@href}%
\providecommand \@href[1]{\@@startlink{#1}\@@href}%
\providecommand \@@href[1]{\endgroup#1\@@endlink}%
\providecommand \@sanitize@url [0]{\catcode `\\12\catcode `\$12\catcode `\&12\catcode `\#12\catcode `\^12\catcode `\_12\catcode `\%12\relax}%
\providecommand \@@startlink[1]{}%
\providecommand \@@endlink[0]{}%
\providecommand \url  [0]{\begingroup\@sanitize@url \@url }%
\providecommand \@url [1]{\endgroup\@href {#1}{\urlprefix }}%
\providecommand \urlprefix  [0]{URL }%
\providecommand \Eprint [0]{\href }%
\providecommand \doibase [0]{http://dx.doi.org/}%
\providecommand \selectlanguage [0]{\@gobble}%
\providecommand \bibinfo  [0]{\@secondoftwo}%
\providecommand \bibfield  [0]{\@secondoftwo}%
\providecommand \translation [1]{[#1]}%
\providecommand \BibitemOpen [0]{}%
\providecommand \bibitemStop [0]{}%
\providecommand \bibitemNoStop [0]{.\EOS\space}%
\providecommand \EOS [0]{\spacefactor3000\relax}%
\providecommand \BibitemShut  [1]{\csname bibitem#1\endcsname}%
\let\auto@bib@innerbib\@empty
\bibitem [{\citenamefont {Raimond}\ \emph {et~al.}(2001)\citenamefont {Raimond}, \citenamefont {Brune},\ and\ \citenamefont {Haroche}}]{RMP2001}%
  \BibitemOpen
  \bibfield  {author} {\bibinfo {author} {\bibfnamefont {J.~M.}\ \bibnamefont {Raimond}}, \bibinfo {author} {\bibfnamefont {M.}~\bibnamefont {Brune}}, \ and\ \bibinfo {author} {\bibfnamefont {S.}~\bibnamefont {Haroche}},\ }\bibfield  {title} {\enquote {\bibinfo {title} {Manipulating quantum entanglement with atoms and photons in a cavity},}\ }\href@noop {} {\bibfield  {journal} {\bibinfo  {journal} {Rev. Mod. Phys.}\ }\textbf {\bibinfo {volume} {73}},\ \bibinfo {pages} {565} (\bibinfo {year} {2001})}\BibitemShut {NoStop}%
\bibitem [{\citenamefont {Reiserer}\ and\ \citenamefont {Rempe}(2015)}]{RMP2015}%
  \BibitemOpen
  \bibfield  {author} {\bibinfo {author} {\bibfnamefont {A.}~\bibnamefont {Reiserer}}\ and\ \bibinfo {author} {\bibfnamefont {G.}~\bibnamefont {Rempe}},\ }\bibfield  {title} {\enquote {\bibinfo {title} {Cavity-based quantum networks with single atoms and optical photons},}\ }\href@noop {} {\bibfield  {journal} {\bibinfo  {journal} {Rev. Mod. Phys.}\ }\textbf {\bibinfo {volume} {87}},\ \bibinfo {pages} {1379} (\bibinfo {year} {2015})}\BibitemShut {NoStop}%
\bibitem [{\citenamefont {Haroche}(2013)}]{RMP2013}%
  \BibitemOpen
  \bibfield  {author} {\bibinfo {author} {\bibfnamefont {S.}~\bibnamefont {Haroche}},\ }\bibfield  {title} {\enquote {\bibinfo {title} {Nobel lecture: Controlling photons in a box and exploring the quantum to classical boundary},}\ }\href@noop {} {\bibfield  {journal} {\bibinfo  {journal} {Rev. Mod. Phys.}\ }\textbf {\bibinfo {volume} {85}},\ \bibinfo {pages} {1083} (\bibinfo {year} {2013})}\BibitemShut {NoStop}%
\bibitem [{\citenamefont {Ritsch}\ \emph {et~al.}(2013)\citenamefont {Ritsch}, \citenamefont {Domokos}, \citenamefont {Brennecke},\ and\ \citenamefont {Esslinger}}]{RMP20132}%
  \BibitemOpen
  \bibfield  {author} {\bibinfo {author} {\bibfnamefont {H.}~\bibnamefont {Ritsch}}, \bibinfo {author} {\bibfnamefont {P.}~\bibnamefont {Domokos}}, \bibinfo {author} {\bibfnamefont {F.}~\bibnamefont {Brennecke}}, \ and\ \bibinfo {author} {\bibfnamefont {T.}~\bibnamefont {Esslinger}},\ }\bibfield  {title} {\enquote {\bibinfo {title} {Cold atoms in cavity-generated dynamical optical potentials},}\ }\href@noop {} {\bibfield  {journal} {\bibinfo  {journal} {Rev. Mod. Phys.}\ }\textbf {\bibinfo {volume} {85}},\ \bibinfo {pages} {553} (\bibinfo {year} {2013})}\BibitemShut {NoStop}%
\bibitem [{\citenamefont {Blais}\ \emph {et~al.}(2021)\citenamefont {Blais}, \citenamefont {Grimsmo}, \citenamefont {Girvin},\ and\ \citenamefont {Wallraff}}]{RMP2021}%
  \BibitemOpen
  \bibfield  {author} {\bibinfo {author} {\bibfnamefont {A.}~\bibnamefont {Blais}}, \bibinfo {author} {\bibfnamefont {A.~L.}\ \bibnamefont {Grimsmo}}, \bibinfo {author} {\bibfnamefont {S.~M.}\ \bibnamefont {Girvin}}, \ and\ \bibinfo {author} {\bibfnamefont {A.}~\bibnamefont {Wallraff}},\ }\bibfield  {title} {\enquote {\bibinfo {title} {Circuit quantum electrodynamics},}\ }\href@noop {} {\bibfield  {journal} {\bibinfo  {journal} {Rev. Mod. Phys.}\ }\textbf {\bibinfo {volume} {93}},\ \bibinfo {pages} {025005} (\bibinfo {year} {2021})}\BibitemShut {NoStop}%
\bibitem [{\citenamefont {McKeever}\ \emph {et~al.}(2003)\citenamefont {McKeever}, \citenamefont {Boca}, \citenamefont {Boozer}, \citenamefont {Buck},\ and\ \citenamefont {Kimble}}]{Nature2003}%
  \BibitemOpen
  \bibfield  {author} {\bibinfo {author} {\bibfnamefont {J.}~\bibnamefont {McKeever}}, \bibinfo {author} {\bibfnamefont {A.}~\bibnamefont {Boca}}, \bibinfo {author} {\bibfnamefont {A.~D.}\ \bibnamefont {Boozer}}, \bibinfo {author} {\bibfnamefont {J.~R.}\ \bibnamefont {Buck}}, \ and\ \bibinfo {author} {\bibfnamefont {H.~J.}\ \bibnamefont {Kimble}},\ }\bibfield  {title} {\enquote {\bibinfo {title} {Experimental realization of a one-atom laser in the regime of strong coupling},}\ }\href@noop {} {\bibfield  {journal} {\bibinfo  {journal} {Nature}\ }\textbf {\bibinfo {volume} {425}},\ \bibinfo {pages} {268} (\bibinfo {year} {2003})}\BibitemShut {NoStop}%
\bibitem [{\citenamefont {McKeever}\ \emph {et~al.}(2004)\citenamefont {McKeever}, \citenamefont {Boca}, \citenamefont {Boozer}, \citenamefont {Miller}, \citenamefont {Buck}, \citenamefont {Kuzmich},\ and\ \citenamefont {Kimble}}]{science.2004}%
  \BibitemOpen
  \bibfield  {author} {\bibinfo {author} {\bibfnamefont {J.}~\bibnamefont {McKeever}}, \bibinfo {author} {\bibfnamefont {A.}~\bibnamefont {Boca}}, \bibinfo {author} {\bibfnamefont {A.~D.}\ \bibnamefont {Boozer}}, \bibinfo {author} {\bibfnamefont {R.}~\bibnamefont {Miller}}, \bibinfo {author} {\bibfnamefont {J.~R.}\ \bibnamefont {Buck}}, \bibinfo {author} {\bibfnamefont {A.}~\bibnamefont {Kuzmich}}, \ and\ \bibinfo {author} {\bibfnamefont {H.~J.}\ \bibnamefont {Kimble}},\ }\bibfield  {title} {\enquote {\bibinfo {title} {Deterministic generation of single photons from one atom trapped in a cavity},}\ }\href@noop {} {\bibfield  {journal} {\bibinfo  {journal} {Science}\ }\textbf {\bibinfo {volume} {303}},\ \bibinfo {pages} {1992} (\bibinfo {year} {2004})}\BibitemShut {NoStop}%
\bibitem [{\citenamefont {Kuhn}\ \emph {et~al.}(2002)\citenamefont {Kuhn}, \citenamefont {Hennrich},\ and\ \citenamefont {Rempe}}]{PRL20022}%
  \BibitemOpen
  \bibfield  {author} {\bibinfo {author} {\bibfnamefont {A.}~\bibnamefont {Kuhn}}, \bibinfo {author} {\bibfnamefont {M.}~\bibnamefont {Hennrich}}, \ and\ \bibinfo {author} {\bibfnamefont {G.}~\bibnamefont {Rempe}},\ }\bibfield  {title} {\enquote {\bibinfo {title} {Deterministic single-photon source for distributed quantum networking},}\ }\href@noop {} {\bibfield  {journal} {\bibinfo  {journal} {Phys. Rev. Lett.}\ }\textbf {\bibinfo {volume} {89}},\ \bibinfo {pages} {067901} (\bibinfo {year} {2002})}\BibitemShut {NoStop}%
\bibitem [{\citenamefont {Morin}\ \emph {et~al.}(2019)\citenamefont {Morin}, \citenamefont {Korber}, \citenamefont {Langenfeld},\ and\ \citenamefont {Rempe}}]{PRL2019}%
  \BibitemOpen
  \bibfield  {author} {\bibinfo {author} {\bibfnamefont {O.}~\bibnamefont {Morin}}, \bibinfo {author} {\bibfnamefont {M.}~\bibnamefont {Korber}}, \bibinfo {author} {\bibfnamefont {S.}~\bibnamefont {Langenfeld}}, \ and\ \bibinfo {author} {\bibfnamefont {G.}~\bibnamefont {Rempe}},\ }\bibfield  {title} {\enquote {\bibinfo {title} {Deterministic shaping and reshaping of single-photon temporal wave functions},}\ }\href@noop {} {\bibfield  {journal} {\bibinfo  {journal} {Phys. Rev. Lett.}\ }\textbf {\bibinfo {volume} {123}},\ \bibinfo {pages} {133602} (\bibinfo {year} {2019})}\BibitemShut {NoStop}%
\bibitem [{\citenamefont {Rosenblum}\ \emph {et~al.}(2016)\citenamefont {Rosenblum}, \citenamefont {Bechler}, \citenamefont {Shomroni}, \citenamefont {Lovsky}, \citenamefont {Guendelman},\ and\ \citenamefont {Dayan}}]{NP2016}%
  \BibitemOpen
  \bibfield  {author} {\bibinfo {author} {\bibfnamefont {S.}~\bibnamefont {Rosenblum}}, \bibinfo {author} {\bibfnamefont {O.}~\bibnamefont {Bechler}}, \bibinfo {author} {\bibfnamefont {I.}~\bibnamefont {Shomroni}}, \bibinfo {author} {\bibfnamefont {Y.}~\bibnamefont {Lovsky}}, \bibinfo {author} {\bibfnamefont {G.}~\bibnamefont {Guendelman}}, \ and\ \bibinfo {author} {\bibfnamefont {B.}~\bibnamefont {Dayan}},\ }\bibfield  {title} {\enquote {\bibinfo {title} {Extraction of a single photon from an optical pulse},}\ }\href@noop {} {\bibfield  {journal} {\bibinfo  {journal} {Nat. Photonics}\ }\textbf {\bibinfo {volume} {10}},\ \bibinfo {pages} {19} (\bibinfo {year} {2016})}\BibitemShut {NoStop}%
\bibitem [{\citenamefont {Hamsen}\ \emph {et~al.}(2017)\citenamefont {Hamsen}, \citenamefont {Tolazzi}, \citenamefont {Wilk},\ and\ \citenamefont {Rempe}}]{PRL2017}%
  \BibitemOpen
  \bibfield  {author} {\bibinfo {author} {\bibfnamefont {C.}~\bibnamefont {Hamsen}}, \bibinfo {author} {\bibfnamefont {K.~N.}\ \bibnamefont {Tolazzi}}, \bibinfo {author} {\bibfnamefont {T.}~\bibnamefont {Wilk}}, \ and\ \bibinfo {author} {\bibfnamefont {G.}~\bibnamefont {Rempe}},\ }\bibfield  {title} {\enquote {\bibinfo {title} {Two-photon blockade in an atom-driven cavity {QED} system},}\ }\href@noop {} {\bibfield  {journal} {\bibinfo  {journal} {Phys. Rev. Lett.}\ }\textbf {\bibinfo {volume} {118}},\ \bibinfo {pages} {133604} (\bibinfo {year} {2017})}\BibitemShut {NoStop}%
\bibitem [{\citenamefont {Schupp}\ \emph {et~al.}(2021)\citenamefont {Schupp}, \citenamefont {Krcmarsky}, \citenamefont {Krutyanskiy}, \citenamefont {Meraner}, \citenamefont {Northup},\ and\ \citenamefont {Lanyon}}]{PRX2021}%
  \BibitemOpen
  \bibfield  {author} {\bibinfo {author} {\bibfnamefont {J.}~\bibnamefont {Schupp}}, \bibinfo {author} {\bibfnamefont {V.}~\bibnamefont {Krcmarsky}}, \bibinfo {author} {\bibfnamefont {V.}~\bibnamefont {Krutyanskiy}}, \bibinfo {author} {\bibfnamefont {M.}~\bibnamefont {Meraner}}, \bibinfo {author} {\bibfnamefont {T.~E.}\ \bibnamefont {Northup}}, \ and\ \bibinfo {author} {\bibfnamefont {B.~P.}\ \bibnamefont {Lanyon}},\ }\bibfield  {title} {\enquote {\bibinfo {title} {Interface between trapped-ion qubits and traveling photons with close-to-optimal efficiency},}\ }\href@noop {} {\bibfield  {journal} {\bibinfo  {journal} {PRX Quantum}\ }\textbf {\bibinfo {volume} {2}},\ \bibinfo {pages} {020331} (\bibinfo {year} {2021})}\BibitemShut {NoStop}%
\bibitem [{\citenamefont {Shomroni}\ \emph {et~al.}(2014)\citenamefont {Shomroni}, \citenamefont {Rosenblum}, \citenamefont {Lovsky}, \citenamefont {Bechler}, \citenamefont {Guendelman},\ and\ \citenamefont {Dayan}}]{Science2014}%
  \BibitemOpen
  \bibfield  {author} {\bibinfo {author} {\bibfnamefont {I.}~\bibnamefont {Shomroni}}, \bibinfo {author} {\bibfnamefont {S.}~\bibnamefont {Rosenblum}}, \bibinfo {author} {\bibfnamefont {Y.}~\bibnamefont {Lovsky}}, \bibinfo {author} {\bibfnamefont {O.}~\bibnamefont {Bechler}}, \bibinfo {author} {\bibfnamefont {G.}~\bibnamefont {Guendelman}}, \ and\ \bibinfo {author} {\bibfnamefont {B.}~\bibnamefont {Dayan}},\ }\bibfield  {title} {\enquote {\bibinfo {title} {All-optical routing of single photons by a one-atom switch controlled by a single photon},}\ }\href@noop {} {\bibfield  {journal} {\bibinfo  {journal} {Science}\ }\textbf {\bibinfo {volume} {345}},\ \bibinfo {pages} {903} (\bibinfo {year} {2014})}\BibitemShut {NoStop}%
\bibitem [{\citenamefont {Daiss}\ \emph {et~al.}(2021)\citenamefont {Daiss}, \citenamefont {Langenfeld}, \citenamefont {Welte}, \citenamefont {Distante}, \citenamefont {Thomas}, \citenamefont {Hartung}, \citenamefont {Morin},\ and\ \citenamefont {Rempe}}]{Science2020}%
  \BibitemOpen
  \bibfield  {author} {\bibinfo {author} {\bibfnamefont {S.}~\bibnamefont {Daiss}}, \bibinfo {author} {\bibfnamefont {S.}~\bibnamefont {Langenfeld}}, \bibinfo {author} {\bibfnamefont {S.}~\bibnamefont {Welte}}, \bibinfo {author} {\bibfnamefont {E.}~\bibnamefont {Distante}}, \bibinfo {author} {\bibfnamefont {P.}~\bibnamefont {Thomas}}, \bibinfo {author} {\bibfnamefont {L.}~\bibnamefont {Hartung}}, \bibinfo {author} {\bibfnamefont {O.}~\bibnamefont {Morin}}, \ and\ \bibinfo {author} {\bibfnamefont {G.}~\bibnamefont {Rempe}},\ }\bibfield  {title} {\enquote {\bibinfo {title} {A quantum-logic gate between distant quantum-network modules},}\ }\href@noop {} {\bibfield  {journal} {\bibinfo  {journal} {Science}\ }\textbf {\bibinfo {volume} {371}},\ \bibinfo {pages} {614} (\bibinfo {year} {2021})}\BibitemShut {NoStop}%
\bibitem [{\citenamefont {Niemietz}\ \emph {et~al.}(2021)\citenamefont {Niemietz}, \citenamefont {Farrera}, \citenamefont {Langenfeld},\ and\ \citenamefont {Rempe}}]{Nature2021}%
  \BibitemOpen
  \bibfield  {author} {\bibinfo {author} {\bibfnamefont {D.}~\bibnamefont {Niemietz}}, \bibinfo {author} {\bibfnamefont {P.}~\bibnamefont {Farrera}}, \bibinfo {author} {\bibfnamefont {S.}~\bibnamefont {Langenfeld}}, \ and\ \bibinfo {author} {\bibfnamefont {G.}~\bibnamefont {Rempe}},\ }\bibfield  {title} {\enquote {\bibinfo {title} {Nondestructive detection of photonic qubits},}\ }\href@noop {} {\bibfield  {journal} {\bibinfo  {journal} {Nature}\ }\textbf {\bibinfo {volume} {591}},\ \bibinfo {pages} {570} (\bibinfo {year} {2021})}\BibitemShut {NoStop}%
\bibitem [{\citenamefont {Saffman}\ \emph {et~al.}(2010)\citenamefont {Saffman}, \citenamefont {Walker},\ and\ \citenamefont {M\o{}lmer}}]{RMP2010}%
  \BibitemOpen
  \bibfield  {author} {\bibinfo {author} {\bibfnamefont {M.}~\bibnamefont {Saffman}}, \bibinfo {author} {\bibfnamefont {T.~G.}\ \bibnamefont {Walker}}, \ and\ \bibinfo {author} {\bibfnamefont {K.}~\bibnamefont {M\o{}lmer}},\ }\bibfield  {title} {\enquote {\bibinfo {title} {Quantum information with rydberg atoms},}\ }\href@noop {} {\bibfield  {journal} {\bibinfo  {journal} {Rev. Mod. Phys.}\ }\textbf {\bibinfo {volume} {82}},\ \bibinfo {pages} {2313} (\bibinfo {year} {2010})}\BibitemShut {NoStop}%
\bibitem [{\citenamefont {Madjarov}\ \emph {et~al.}(2020)\citenamefont {Madjarov}, \citenamefont {Covey}, \citenamefont {Shaw}, \citenamefont {Choi}, \citenamefont {Kale}, \citenamefont {Cooper}, \citenamefont {Pichler}, \citenamefont {Schkolnik}, \citenamefont {Williams},\ and\ \citenamefont {Endres}}]{NP2020}%
  \BibitemOpen
  \bibfield  {author} {\bibinfo {author} {\bibfnamefont {I.~S.}\ \bibnamefont {Madjarov}}, \bibinfo {author} {\bibfnamefont {J.~P.}\ \bibnamefont {Covey}}, \bibinfo {author} {\bibfnamefont {A.~L.}\ \bibnamefont {Shaw}}, \bibinfo {author} {\bibfnamefont {J.}~\bibnamefont {Choi}}, \bibinfo {author} {\bibfnamefont {A.}~\bibnamefont {Kale}}, \bibinfo {author} {\bibfnamefont {A.}~\bibnamefont {Cooper}}, \bibinfo {author} {\bibfnamefont {H.}~\bibnamefont {Pichler}}, \bibinfo {author} {\bibfnamefont {V.}~\bibnamefont {Schkolnik}}, \bibinfo {author} {\bibfnamefont {J.~R.}\ \bibnamefont {Williams}}, \ and\ \bibinfo {author} {\bibfnamefont {M.}~\bibnamefont {Endres}},\ }\bibfield  {title} {\enquote {\bibinfo {title} {High-fidelity entanglement and detection of alkaline-earth {R}ydberg atoms},}\ }\href@noop {} {\bibfield  {journal} {\bibinfo  {journal} {Nat. Phys.}\ }\textbf {\bibinfo {volume} {16}},\ \bibinfo {pages} {857} (\bibinfo {year} {2020})}\BibitemShut {NoStop}%
\bibitem [{\citenamefont {Graham}\ \emph {et~al.}(2022)\citenamefont {Graham}, \citenamefont {Song}, \citenamefont {Scott}, \citenamefont {Poole}, \citenamefont {Phuttitarn}, \citenamefont {Jooya}, \citenamefont {Eichler}, \citenamefont {Jiang}, \citenamefont {Marra}, \citenamefont {Grinkemeyer}, \citenamefont {Kwon}, \citenamefont {Ebert}, \citenamefont {Cherek}, \citenamefont {Lichtman}, \citenamefont {Gillette}, \citenamefont {Gilbert}, \citenamefont {Bowman}, \citenamefont {Ballance}, \citenamefont {Campbell}, \citenamefont {Dahl}, \citenamefont {Crawford}, \citenamefont {Blunt}, \citenamefont {Noel},\ and\ \citenamefont {Saffman}}]{Nature20221}%
  \BibitemOpen
  \bibfield  {author} {\bibinfo {author} {\bibfnamefont {T.~M.}\ \bibnamefont {Graham}}, \bibinfo {author} {\bibfnamefont {Y.}~\bibnamefont {Song}}, \bibinfo {author} {\bibfnamefont {J.~A.}\ \bibnamefont {Scott}}, \bibinfo {author} {\bibfnamefont {C.}~\bibnamefont {Poole}}, \bibinfo {author} {\bibfnamefont {L.}~\bibnamefont {Phuttitarn}}, \bibinfo {author} {\bibfnamefont {K.}~\bibnamefont {Jooya}}, \bibinfo {author} {\bibfnamefont {P.}~\bibnamefont {Eichler}}, \bibinfo {author} {\bibfnamefont {X.}~\bibnamefont {Jiang}}, \bibinfo {author} {\bibfnamefont {A.~A.}\ \bibnamefont {Marra}}, \bibinfo {author} {\bibfnamefont {B.}~\bibnamefont {Grinkemeyer}}, \bibinfo {author} {\bibfnamefont {M.}~\bibnamefont {Kwon}}, \bibinfo {author} {\bibfnamefont {M.}~\bibnamefont {Ebert}}, \bibinfo {author} {\bibfnamefont {J.}~\bibnamefont {Cherek}}, \bibinfo {author} {\bibfnamefont {M.}~\bibnamefont {Lichtman}}, \bibinfo {author} {\bibfnamefont {M.}~\bibnamefont {Gillette}}, \bibinfo {author} {\bibfnamefont {J.~P.}\ \bibnamefont
  {Gilbert}}, \bibinfo {author} {\bibfnamefont {D.~N.}\ \bibnamefont {Bowman}}, \bibinfo {author} {\bibfnamefont {T.}~\bibnamefont {Ballance}}, \bibinfo {author} {\bibfnamefont {C.}~\bibnamefont {Campbell}}, \bibinfo {author} {\bibfnamefont {E.~D.}\ \bibnamefont {Dahl}}, \bibinfo {author} {\bibfnamefont {O.}~\bibnamefont {Crawford}}, \bibinfo {author} {\bibfnamefont {N.}~\bibnamefont {Blunt}}, \bibinfo {author} {\bibfnamefont {B.~R.}\ \bibnamefont {Noel}}, \ and\ \bibinfo {author} {\bibfnamefont {M.}~\bibnamefont {Saffman}},\ }\bibfield  {title} {\enquote {\bibinfo {title} {Multi-qubit entanglement and algorithms on a neutral-atom quantum computer},}\ }\href@noop {} {\bibfield  {journal} {\bibinfo  {journal} {Nature}\ }\textbf {\bibinfo {volume} {604}},\ \bibinfo {pages} {457} (\bibinfo {year} {2022})}\BibitemShut {NoStop}%
\bibitem [{\citenamefont {Barredo}\ \emph {et~al.}(2018)\citenamefont {Barredo}, \citenamefont {Lienhard}, \citenamefont {de~Léséleuc}, \citenamefont {Lahaye},\ and\ \citenamefont {Browaeys}}]{Nature2018}%
  \BibitemOpen
  \bibfield  {author} {\bibinfo {author} {\bibfnamefont {D.}~\bibnamefont {Barredo}}, \bibinfo {author} {\bibfnamefont {V.}~\bibnamefont {Lienhard}}, \bibinfo {author} {\bibfnamefont {S.}~\bibnamefont {de~Léséleuc}}, \bibinfo {author} {\bibfnamefont {T.}~\bibnamefont {Lahaye}}, \ and\ \bibinfo {author} {\bibfnamefont {A.}~\bibnamefont {Browaeys}},\ }\bibfield  {title} {\enquote {\bibinfo {title} {Synthetic three-dimensional atomic structures assembled atom by atom},}\ }\href@noop {} {\bibfield  {journal} {\bibinfo  {journal} {Nature}\ }\textbf {\bibinfo {volume} {561}},\ \bibinfo {pages} {79} (\bibinfo {year} {2018})}\BibitemShut {NoStop}%
\bibitem [{\citenamefont {Bluvstein}\ \emph {et~al.}(2022)\citenamefont {Bluvstein}, \citenamefont {Levine}, \citenamefont {Semeghini}, \citenamefont {Wang}, \citenamefont {Ebadi}, \citenamefont {Kalinowski}, \citenamefont {Keesling}, \citenamefont {Maskara}, \citenamefont {Pichler}, \citenamefont {Greiner}, \citenamefont {Vuletić},\ and\ \citenamefont {Lukin}}]{Nature2022}%
  \BibitemOpen
  \bibfield  {author} {\bibinfo {author} {\bibfnamefont {D.}~\bibnamefont {Bluvstein}}, \bibinfo {author} {\bibfnamefont {H.}~\bibnamefont {Levine}}, \bibinfo {author} {\bibfnamefont {G.}~\bibnamefont {Semeghini}}, \bibinfo {author} {\bibfnamefont {T.~T.}\ \bibnamefont {Wang}}, \bibinfo {author} {\bibfnamefont {S.}~\bibnamefont {Ebadi}}, \bibinfo {author} {\bibfnamefont {M.}~\bibnamefont {Kalinowski}}, \bibinfo {author} {\bibfnamefont {A.}~\bibnamefont {Keesling}}, \bibinfo {author} {\bibfnamefont {N.}~\bibnamefont {Maskara}}, \bibinfo {author} {\bibfnamefont {H.}~\bibnamefont {Pichler}}, \bibinfo {author} {\bibfnamefont {M.}~\bibnamefont {Greiner}}, \bibinfo {author} {\bibfnamefont {V.}~\bibnamefont {Vuletić}}, \ and\ \bibinfo {author} {\bibfnamefont {M.~D.}\ \bibnamefont {Lukin}},\ }\bibfield  {title} {\enquote {\bibinfo {title} {A quantum processor based on coherent transport of entangled atom arrays},}\ }\href@noop {} {\bibfield  {journal} {\bibinfo  {journal} {Nature}\ }\textbf {\bibinfo {volume}
  {604}},\ \bibinfo {pages} {451} (\bibinfo {year} {2022})}\BibitemShut {NoStop}%
\bibitem [{\citenamefont {Evered}\ \emph {et~al.}(2023)\citenamefont {Evered}, \citenamefont {Bluvstein}, \citenamefont {Kalinowski}, \citenamefont {Ebadi}, \citenamefont {Manovitz}, \citenamefont {Zhou}, \citenamefont {Li}, \citenamefont {Geim}, \citenamefont {Wang}, \citenamefont {Maskara}, \citenamefont {Levine}, \citenamefont {Semeghini}, \citenamefont {Greiner}, \citenamefont {Vuletić},\ and\ \citenamefont {Lukin}}]{Nature2023}%
  \BibitemOpen
  \bibfield  {author} {\bibinfo {author} {\bibfnamefont {S.~J.}\ \bibnamefont {Evered}}, \bibinfo {author} {\bibfnamefont {D.}~\bibnamefont {Bluvstein}}, \bibinfo {author} {\bibfnamefont {M.}~\bibnamefont {Kalinowski}}, \bibinfo {author} {\bibfnamefont {S.}~\bibnamefont {Ebadi}}, \bibinfo {author} {\bibfnamefont {T.}~\bibnamefont {Manovitz}}, \bibinfo {author} {\bibfnamefont {H.}~\bibnamefont {Zhou}}, \bibinfo {author} {\bibfnamefont {S.~H.}\ \bibnamefont {Li}}, \bibinfo {author} {\bibfnamefont {A.~A.}\ \bibnamefont {Geim}}, \bibinfo {author} {\bibfnamefont {T.~T.}\ \bibnamefont {Wang}}, \bibinfo {author} {\bibfnamefont {N.}~\bibnamefont {Maskara}}, \bibinfo {author} {\bibfnamefont {H.}~\bibnamefont {Levine}}, \bibinfo {author} {\bibfnamefont {G.}~\bibnamefont {Semeghini}}, \bibinfo {author} {\bibfnamefont {M.}~\bibnamefont {Greiner}}, \bibinfo {author} {\bibfnamefont {V.}~\bibnamefont {Vuletić}}, \ and\ \bibinfo {author} {\bibfnamefont {M.~D.}\ \bibnamefont {Lukin}},\ }\bibfield  {title} {\enquote {\bibinfo
  {title} {High-fidelity parallel entangling gates on a neutral-atom quantum computer},}\ }\href@noop {} {\bibfield  {journal} {\bibinfo  {journal} {Nature}\ }\textbf {\bibinfo {volume} {622}},\ \bibinfo {pages} {268} (\bibinfo {year} {2023})}\BibitemShut {NoStop}%
\bibitem [{\citenamefont {Bluvstein}\ \emph {et~al.}(2024)\citenamefont {Bluvstein}, \citenamefont {Evered}, \citenamefont {Geim}, \citenamefont {Li}, \citenamefont {Zhou}, \citenamefont {Manovitz}, \citenamefont {Ebadi}, \citenamefont {Cain}, \citenamefont {Kalinowski}, \citenamefont {Hangleiter}, \citenamefont {Ataides}, \citenamefont {Maskara}, \citenamefont {Cong}, \citenamefont {Gao}, \citenamefont {Rodriguez}, \citenamefont {Karolyshyn}, \citenamefont {Semeghini}, \citenamefont {Gullans}, \citenamefont {Greiner}, \citenamefont {Vuletić},\ and\ \citenamefont {Lukin}}]{Nature2024}%
  \BibitemOpen
  \bibfield  {author} {\bibinfo {author} {\bibfnamefont {D.}~\bibnamefont {Bluvstein}}, \bibinfo {author} {\bibfnamefont {S.~J.}\ \bibnamefont {Evered}}, \bibinfo {author} {\bibfnamefont {A.~A.}\ \bibnamefont {Geim}}, \bibinfo {author} {\bibfnamefont {S.~H.}\ \bibnamefont {Li}}, \bibinfo {author} {\bibfnamefont {H.}~\bibnamefont {Zhou}}, \bibinfo {author} {\bibfnamefont {T.}~\bibnamefont {Manovitz}}, \bibinfo {author} {\bibfnamefont {S.}~\bibnamefont {Ebadi}}, \bibinfo {author} {\bibfnamefont {M.}~\bibnamefont {Cain}}, \bibinfo {author} {\bibfnamefont {M.}~\bibnamefont {Kalinowski}}, \bibinfo {author} {\bibfnamefont {D.}~\bibnamefont {Hangleiter}}, \bibinfo {author} {\bibfnamefont {J.~P.~B.}\ \bibnamefont {Ataides}}, \bibinfo {author} {\bibfnamefont {N.}~\bibnamefont {Maskara}}, \bibinfo {author} {\bibfnamefont {I.}~\bibnamefont {Cong}}, \bibinfo {author} {\bibfnamefont {X.}~\bibnamefont {Gao}}, \bibinfo {author} {\bibfnamefont {P.~S.}\ \bibnamefont {Rodriguez}}, \bibinfo {author} {\bibfnamefont
  {T.}~\bibnamefont {Karolyshyn}}, \bibinfo {author} {\bibfnamefont {G.}~\bibnamefont {Semeghini}}, \bibinfo {author} {\bibfnamefont {M.~J.}\ \bibnamefont {Gullans}}, \bibinfo {author} {\bibfnamefont {M.}~\bibnamefont {Greiner}}, \bibinfo {author} {\bibfnamefont {V.}~\bibnamefont {Vuletić}}, \ and\ \bibinfo {author} {\bibfnamefont {M.~D.}\ \bibnamefont {Lukin}},\ }\bibfield  {title} {\enquote {\bibinfo {title} {Logical quantum processor based on reconfigurable atom arrays},}\ }\href@noop {} {\bibfield  {journal} {\bibinfo  {journal} {Nature}\ }\textbf {\bibinfo {volume} {626}},\ \bibinfo {pages} {58} (\bibinfo {year} {2024})}\BibitemShut {NoStop}%
\bibitem [{\citenamefont {Liu}\ \emph {et~al.}(2023)\citenamefont {Liu}, \citenamefont {Wang}, \citenamefont {Yang}, \citenamefont {Wang}, \citenamefont {Fan}, \citenamefont {Guan}, \citenamefont {Li}, \citenamefont {Zhang},\ and\ \citenamefont {Zhang}}]{PRL2023}%
  \BibitemOpen
  \bibfield  {author} {\bibinfo {author} {\bibfnamefont {Y.}~\bibnamefont {Liu}}, \bibinfo {author} {\bibfnamefont {Z.}~\bibnamefont {Wang}}, \bibinfo {author} {\bibfnamefont {P.}~\bibnamefont {Yang}}, \bibinfo {author} {\bibfnamefont {Q.}~\bibnamefont {Wang}}, \bibinfo {author} {\bibfnamefont {Q.}~\bibnamefont {Fan}}, \bibinfo {author} {\bibfnamefont {S.}~\bibnamefont {Guan}}, \bibinfo {author} {\bibfnamefont {G.}~\bibnamefont {Li}}, \bibinfo {author} {\bibfnamefont {P.}~\bibnamefont {Zhang}}, \ and\ \bibinfo {author} {\bibfnamefont {T.}~\bibnamefont {Zhang}},\ }\bibfield  {title} {\enquote {\bibinfo {title} {Realization of strong coupling between deterministic single-atom arrays and a high-finesse miniature optical cavity},}\ }\href@noop {} {\bibfield  {journal} {\bibinfo  {journal} {Phys. Rev. Lett.}\ }\textbf {\bibinfo {volume} {130}},\ \bibinfo {pages} {173601} (\bibinfo {year} {2023})}\BibitemShut {NoStop}%
\bibitem [{\citenamefont {Zhang}\ \emph {et~al.}(2024)\citenamefont {Zhang}, \citenamefont {Yu}, \citenamefont {Zhang}, \citenamefont {Xiang},\ and\ \citenamefont {Zhang}}]{PRR2024}%
  \BibitemOpen
  \bibfield  {author} {\bibinfo {author} {\bibfnamefont {X.}~\bibnamefont {Zhang}}, \bibinfo {author} {\bibfnamefont {Z.}~\bibnamefont {Yu}}, \bibinfo {author} {\bibfnamefont {H.}~\bibnamefont {Zhang}}, \bibinfo {author} {\bibfnamefont {D.}~\bibnamefont {Xiang}}, \ and\ \bibinfo {author} {\bibfnamefont {H.}~\bibnamefont {Zhang}},\ }\bibfield  {title} {\enquote {\bibinfo {title} {Cavity dark mode mediated by atom array without atomic scattering loss},}\ }\href@noop {} {\bibfield  {journal} {\bibinfo  {journal} {Phys. Rev. Research}\ }\textbf {\bibinfo {volume} {6}},\ \bibinfo {pages} {042026} (\bibinfo {year} {2024})}\BibitemShut {NoStop}%
\bibitem [{\citenamefont {Yan}\ \emph {et~al.}(2023)\citenamefont {Yan}, \citenamefont {Ho}, \citenamefont {Lu}, \citenamefont {Masson}, \citenamefont {Asenjo-Garcia},\ and\ \citenamefont {Stamper-Kurn}}]{PRL20232}%
  \BibitemOpen
  \bibfield  {author} {\bibinfo {author} {\bibfnamefont {Z.}~\bibnamefont {Yan}}, \bibinfo {author} {\bibfnamefont {J.}~\bibnamefont {Ho}}, \bibinfo {author} {\bibfnamefont {Y.-H.}\ \bibnamefont {Lu}}, \bibinfo {author} {\bibfnamefont {S.~J.}\ \bibnamefont {Masson}}, \bibinfo {author} {\bibfnamefont {A.}~\bibnamefont {Asenjo-Garcia}}, \ and\ \bibinfo {author} {\bibfnamefont {D.~M.}\ \bibnamefont {Stamper-Kurn}},\ }\bibfield  {title} {\enquote {\bibinfo {title} {Superradiant and subradiant cavity scattering by atom arrays},}\ }\href@noop {} {\bibfield  {journal} {\bibinfo  {journal} {Phys. Rev. Lett.}\ }\textbf {\bibinfo {volume} {131}},\ \bibinfo {pages} {253603} (\bibinfo {year} {2023})}\BibitemShut {NoStop}%
\bibitem [{\citenamefont {Ye}\ \emph {et~al.}(2023)\citenamefont {Ye}, \citenamefont {Tian}, \citenamefont {Lin}, \citenamefont {Luo}, \citenamefont {You}, \citenamefont {Hu}, \citenamefont {Zhang}, \citenamefont {Chen},\ and\ \citenamefont {Li}}]{PRL20233}%
  \BibitemOpen
  \bibfield  {author} {\bibinfo {author} {\bibfnamefont {M.}~\bibnamefont {Ye}}, \bibinfo {author} {\bibfnamefont {Y.}~\bibnamefont {Tian}}, \bibinfo {author} {\bibfnamefont {J.}~\bibnamefont {Lin}}, \bibinfo {author} {\bibfnamefont {Y.}~\bibnamefont {Luo}}, \bibinfo {author} {\bibfnamefont {J.}~\bibnamefont {You}}, \bibinfo {author} {\bibfnamefont {J.}~\bibnamefont {Hu}}, \bibinfo {author} {\bibfnamefont {W.}~\bibnamefont {Zhang}}, \bibinfo {author} {\bibfnamefont {W.}~\bibnamefont {Chen}}, \ and\ \bibinfo {author} {\bibfnamefont {X.}~\bibnamefont {Li}},\ }\bibfield  {title} {\enquote {\bibinfo {title} {Universal quantum optimization with cold atoms in an optical cavity},}\ }\href@noop {} {\bibfield  {journal} {\bibinfo  {journal} {Phys. Rev. Lett.}\ }\textbf {\bibinfo {volume} {131}},\ \bibinfo {pages} {103601} (\bibinfo {year} {2023})}\BibitemShut {NoStop}%
\bibitem [{\citenamefont {Ye}\ and\ \citenamefont {Li}(2024)}]{QF2024}%
  \BibitemOpen
  \bibfield  {author} {\bibinfo {author} {\bibfnamefont {M.}~\bibnamefont {Ye}}\ and\ \bibinfo {author} {\bibfnamefont {X.}~\bibnamefont {Li}},\ }\bibfield  {title} {\enquote {\bibinfo {title} {Atom cavity encoding for np-complete problems},}\ }\href@noop {} {\bibfield  {journal} {\bibinfo  {journal} {Quantum Front.}\ }\textbf {\bibinfo {volume} {3}},\ \bibinfo {pages} {24} (\bibinfo {year} {2024})}\BibitemShut {NoStop}%
\bibitem [{\citenamefont {Davis}\ \emph {et~al.}(2019)\citenamefont {Davis}, \citenamefont {Bentsen}, \citenamefont {Homeier}, \citenamefont {Li},\ and\ \citenamefont {Schleier-Smith}}]{PRL20192}%
  \BibitemOpen
  \bibfield  {author} {\bibinfo {author} {\bibfnamefont {E.~J.}\ \bibnamefont {Davis}}, \bibinfo {author} {\bibfnamefont {G.}~\bibnamefont {Bentsen}}, \bibinfo {author} {\bibfnamefont {L.}~\bibnamefont {Homeier}}, \bibinfo {author} {\bibfnamefont {T.}~\bibnamefont {Li}}, \ and\ \bibinfo {author} {\bibfnamefont {M.~H.}\ \bibnamefont {Schleier-Smith}},\ }\bibfield  {title} {\enquote {\bibinfo {title} {Photon-mediated spin-exchange dynamics of spin-1 atoms},}\ }\href@noop {} {\bibfield  {journal} {\bibinfo  {journal} {Phys. Rev. Lett.}\ }\textbf {\bibinfo {volume} {122}},\ \bibinfo {pages} {010405} (\bibinfo {year} {2019})}\BibitemShut {NoStop}%
\bibitem [{\citenamefont {Periwal}\ \emph {et~al.}(2021)\citenamefont {Periwal}, \citenamefont {Cooper}, \citenamefont {Kunkel}, \citenamefont {Wienand}, \citenamefont {Davis},\ and\ \citenamefont {Schleier-Smith}}]{Nature20212}%
  \BibitemOpen
  \bibfield  {author} {\bibinfo {author} {\bibfnamefont {A.}~\bibnamefont {Periwal}}, \bibinfo {author} {\bibfnamefont {E.~S.}\ \bibnamefont {Cooper}}, \bibinfo {author} {\bibfnamefont {P.}~\bibnamefont {Kunkel}}, \bibinfo {author} {\bibfnamefont {J.~F.}\ \bibnamefont {Wienand}}, \bibinfo {author} {\bibfnamefont {E.~J.}\ \bibnamefont {Davis}}, \ and\ \bibinfo {author} {\bibfnamefont {M.}~\bibnamefont {Schleier-Smith}},\ }\bibfield  {title} {\enquote {\bibinfo {title} {Programmable interactions and emergent geometry in an array of atom clouds},}\ }\href@noop {} {\bibfield  {journal} {\bibinfo  {journal} {Nature}\ }\textbf {\bibinfo {volume} {600}},\ \bibinfo {pages} {630} (\bibinfo {year} {2021})}\BibitemShut {NoStop}%
\bibitem [{\citenamefont {Muniz}\ \emph {et~al.}(2020)\citenamefont {Muniz}, \citenamefont {Barberena}, \citenamefont {Lewis-Swan}, \citenamefont {Young}, \citenamefont {Cline}, \citenamefont {Rey},\ and\ \citenamefont {Thompson}}]{Nature2020}%
  \BibitemOpen
  \bibfield  {author} {\bibinfo {author} {\bibfnamefont {J.~A.}\ \bibnamefont {Muniz}}, \bibinfo {author} {\bibfnamefont {D.}~\bibnamefont {Barberena}}, \bibinfo {author} {\bibfnamefont {R.~J.}\ \bibnamefont {Lewis-Swan}}, \bibinfo {author} {\bibfnamefont {D.~J.}\ \bibnamefont {Young}}, \bibinfo {author} {\bibfnamefont {J.~R.~K.}\ \bibnamefont {Cline}}, \bibinfo {author} {\bibfnamefont {A.~M.}\ \bibnamefont {Rey}}, \ and\ \bibinfo {author} {\bibfnamefont {J.~K.}\ \bibnamefont {Thompson}},\ }\bibfield  {title} {\enquote {\bibinfo {title} {Exploring dynamical phase transitions with cold atoms in an optical cavity},}\ }\href@noop {} {\bibfield  {journal} {\bibinfo  {journal} {Nature}\ }\textbf {\bibinfo {volume} {580}},\ \bibinfo {pages} {602} (\bibinfo {year} {2020})}\BibitemShut {NoStop}%
\bibitem [{\citenamefont {Young}\ \emph {et~al.}(2024)\citenamefont {Young}, \citenamefont {Chu}, \citenamefont {Song}, \citenamefont {Barberena}, \citenamefont {Wellnitz}, \citenamefont {Niu}, \citenamefont {Schäfer}, \citenamefont {Lewis-Swan}, \citenamefont {Rey},\ and\ \citenamefont {Thompson}}]{Nature20242}%
  \BibitemOpen
  \bibfield  {author} {\bibinfo {author} {\bibfnamefont {D.~J.}\ \bibnamefont {Young}}, \bibinfo {author} {\bibfnamefont {A.~J.}\ \bibnamefont {Chu}}, \bibinfo {author} {\bibfnamefont {E.~Y.}\ \bibnamefont {Song}}, \bibinfo {author} {\bibfnamefont {D.}~\bibnamefont {Barberena}}, \bibinfo {author} {\bibfnamefont {D.}~\bibnamefont {Wellnitz}}, \bibinfo {author} {\bibfnamefont {Z.~J.}\ \bibnamefont {Niu}}, \bibinfo {author} {\bibfnamefont {V.~M.}\ \bibnamefont {Schäfer}}, \bibinfo {author} {\bibfnamefont {R.~J.}\ \bibnamefont {Lewis-Swan}}, \bibinfo {author} {\bibfnamefont {A.~M.}\ \bibnamefont {Rey}}, \ and\ \bibinfo {author} {\bibfnamefont {J.~K.}\ \bibnamefont {Thompson}},\ }\bibfield  {title} {\enquote {\bibinfo {title} {Observing dynamical phases of bcs superconductors in a cavity {QED} simulator},}\ }\href@noop {} {\bibfield  {journal} {\bibinfo  {journal} {Nature}\ }\textbf {\bibinfo {volume} {625}},\ \bibinfo {pages} {679} (\bibinfo {year} {2024})}\BibitemShut {NoStop}%
\bibitem [{\citenamefont {Hartung}\ \emph {et~al.}(2024)\citenamefont {Hartung}, \citenamefont {Seubert}, \citenamefont {Welte}, \citenamefont {Distante1},\ and\ \citenamefont {Rempe}}]{Science2024}%
  \BibitemOpen
  \bibfield  {author} {\bibinfo {author} {\bibfnamefont {L.}~\bibnamefont {Hartung}}, \bibinfo {author} {\bibfnamefont {M.}~\bibnamefont {Seubert}}, \bibinfo {author} {\bibfnamefont {S.}~\bibnamefont {Welte}}, \bibinfo {author} {\bibfnamefont {E.}~\bibnamefont {Distante1}}, \ and\ \bibinfo {author} {\bibfnamefont {G.}~\bibnamefont {Rempe}},\ }\bibfield  {title} {\enquote {\bibinfo {title} {A quantum-network register assembled with optical tweezers in an optical cavity},}\ }\href@noop {} {\bibfield  {journal} {\bibinfo  {journal} {Science}\ }\textbf {\bibinfo {volume} {385}},\ \bibinfo {pages} {179} (\bibinfo {year} {2024})}\BibitemShut {NoStop}%
\bibitem [{\citenamefont {Deist}\ \emph {et~al.}(2022)\citenamefont {Deist}, \citenamefont {Lu}, \citenamefont {Ho}, \citenamefont {Pasha}, \citenamefont {Zeiher}, \citenamefont {Yan},\ and\ \citenamefont {Stamper-Kurn}}]{PRL2022}%
  \BibitemOpen
  \bibfield  {author} {\bibinfo {author} {\bibfnamefont {E.}~\bibnamefont {Deist}}, \bibinfo {author} {\bibfnamefont {Y.~H.}\ \bibnamefont {Lu}}, \bibinfo {author} {\bibfnamefont {J.}~\bibnamefont {Ho}}, \bibinfo {author} {\bibfnamefont {M.~K.}\ \bibnamefont {Pasha}}, \bibinfo {author} {\bibfnamefont {J.}~\bibnamefont {Zeiher}}, \bibinfo {author} {\bibfnamefont {Z.}~\bibnamefont {Yan}}, \ and\ \bibinfo {author} {\bibfnamefont {D.~M.}\ \bibnamefont {Stamper-Kurn}},\ }\bibfield  {title} {\enquote {\bibinfo {title} {Mid-circuit cavity measurement in a neutral atom array},}\ }\href@noop {} {\bibfield  {journal} {\bibinfo  {journal} {Phys. Rev. Lett.}\ }\textbf {\bibinfo {volume} {129}},\ \bibinfo {pages} {203602} (\bibinfo {year} {2022})}\BibitemShut {NoStop}%
\bibitem [{\citenamefont {Reiserer}\ \emph {et~al.}(2013)\citenamefont {Reiserer}, \citenamefont {Nolleke}, \citenamefont {Ritter},\ and\ \citenamefont {Rempe}}]{PRL2013}%
  \BibitemOpen
  \bibfield  {author} {\bibinfo {author} {\bibfnamefont {A.}~\bibnamefont {Reiserer}}, \bibinfo {author} {\bibfnamefont {C.}~\bibnamefont {Nolleke}}, \bibinfo {author} {\bibfnamefont {S.}~\bibnamefont {Ritter}}, \ and\ \bibinfo {author} {\bibfnamefont {G.}~\bibnamefont {Rempe}},\ }\bibfield  {title} {\enquote {\bibinfo {title} {Ground-state cooling of a single atom at the center of an optical cavity},}\ }\href@noop {} {\bibfield  {journal} {\bibinfo  {journal} {Phys. Rev. Lett.}\ }\textbf {\bibinfo {volume} {110}},\ \bibinfo {pages} {223003} (\bibinfo {year} {2013})}\BibitemShut {NoStop}%
\bibitem [{\citenamefont {Li}\ \emph {et~al.}(2020)\citenamefont {Li}, \citenamefont {Li}, \citenamefont {Wu}, \citenamefont {Fan}, \citenamefont {Yang}, \citenamefont {Zhang},\ and\ \citenamefont {Zhang}}]{RSI2020}%
  \BibitemOpen
  \bibfield  {author} {\bibinfo {author} {\bibfnamefont {S.}~\bibnamefont {Li}}, \bibinfo {author} {\bibfnamefont {G.}~\bibnamefont {Li}}, \bibinfo {author} {\bibfnamefont {W.}~\bibnamefont {Wu}}, \bibinfo {author} {\bibfnamefont {Q.}~\bibnamefont {Fan}}, \bibinfo {author} {\bibfnamefont {P.}~\bibnamefont {Yang}}, \bibinfo {author} {\bibfnamefont {P.}~\bibnamefont {Zhang}}, \ and\ \bibinfo {author} {\bibfnamefont {T.}~\bibnamefont {Zhang}},\ }\bibfield  {title} {\enquote {\bibinfo {title} {High-numerical-aperture and long-working-distance objective for single-atom experiments},}\ }\href@noop {} {\bibfield  {journal} {\bibinfo  {journal} {Rev. Sci. Instrum.}\ }\textbf {\bibinfo {volume} {91}},\ \bibinfo {pages} {043104} (\bibinfo {year} {2020})}\BibitemShut {NoStop}%
\bibitem [{\citenamefont {Schlosser}\ \emph {et~al.}(2002)\citenamefont {Schlosser}, \citenamefont {Reymond},\ and\ \citenamefont {Grangier}}]{PRL2002}%
  \BibitemOpen
  \bibfield  {author} {\bibinfo {author} {\bibfnamefont {N.}~\bibnamefont {Schlosser}}, \bibinfo {author} {\bibfnamefont {G.}~\bibnamefont {Reymond}}, \ and\ \bibinfo {author} {\bibfnamefont {P.}~\bibnamefont {Grangier}},\ }\bibfield  {title} {\enquote {\bibinfo {title} {Collisional blockade in microscopic optical dipole traps},}\ }\href@noop {} {\bibfield  {journal} {\bibinfo  {journal} {Phys. Rev. Lett.}\ }\textbf {\bibinfo {volume} {89}},\ \bibinfo {pages} {023005} (\bibinfo {year} {2002})}\BibitemShut {NoStop}%
\bibitem [{\citenamefont {Endres}\ \emph {et~al.}(2016)\citenamefont {Endres}, \citenamefont {Bernien}, \citenamefont {Keesling}, \citenamefont {Levine}, \citenamefont {Anschuetz}, \citenamefont {Krajenbrink}, \citenamefont {Senko}, \citenamefont {Vuletic}, \citenamefont {Greiner},\ and\ \citenamefont {Lukin}}]{Science2016}%
  \BibitemOpen
  \bibfield  {author} {\bibinfo {author} {\bibfnamefont {M.}~\bibnamefont {Endres}}, \bibinfo {author} {\bibfnamefont {H.}~\bibnamefont {Bernien}}, \bibinfo {author} {\bibfnamefont {A.}~\bibnamefont {Keesling}}, \bibinfo {author} {\bibfnamefont {H.}~\bibnamefont {Levine}}, \bibinfo {author} {\bibfnamefont {E.~R.}\ \bibnamefont {Anschuetz}}, \bibinfo {author} {\bibfnamefont {A.}~\bibnamefont {Krajenbrink}}, \bibinfo {author} {\bibfnamefont {C.}~\bibnamefont {Senko}}, \bibinfo {author} {\bibfnamefont {V.}~\bibnamefont {Vuletic}}, \bibinfo {author} {\bibfnamefont {M.}~\bibnamefont {Greiner}}, \ and\ \bibinfo {author} {\bibfnamefont {M.~D.}\ \bibnamefont {Lukin}},\ }\bibfield  {title} {\enquote {\bibinfo {title} {Atom-by-atom assembly of defect-free one-dimensional cold atom arrays},}\ }\href@noop {} {\bibfield  {journal} {\bibinfo  {journal} {Science}\ }\textbf {\bibinfo {volume} {354}},\ \bibinfo {pages} {1024} (\bibinfo {year} {2016})}\BibitemShut {NoStop}%
\bibitem [{\citenamefont {Dudin}\ \emph {et~al.}(2012)\citenamefont {Dudin}, \citenamefont {Li}, \citenamefont {Bariani},\ and\ \citenamefont {Kuzmich}}]{NP2012}%
  \BibitemOpen
  \bibfield  {author} {\bibinfo {author} {\bibfnamefont {Y.~O.}\ \bibnamefont {Dudin}}, \bibinfo {author} {\bibfnamefont {L.}~\bibnamefont {Li}}, \bibinfo {author} {\bibfnamefont {F.}~\bibnamefont {Bariani}}, \ and\ \bibinfo {author} {\bibfnamefont {A.}~\bibnamefont {Kuzmich}},\ }\bibfield  {title} {\enquote {\bibinfo {title} {Observation of coherent many-body rabi oscillations},}\ }\href@noop {} {\bibfield  {journal} {\bibinfo  {journal} {Nat. Phys.}\ }\textbf {\bibinfo {volume} {8}},\ \bibinfo {pages} {790} (\bibinfo {year} {2012})}\BibitemShut {NoStop}%
\bibitem [{\citenamefont {Labuhn}\ \emph {et~al.}(2016)\citenamefont {Labuhn}, \citenamefont {Barredo}, \citenamefont {Ravets}, \citenamefont {de~Leseleuc}, \citenamefont {Macri}, \citenamefont {Lahaye},\ and\ \citenamefont {Browaeys}}]{Nature2016}%
  \BibitemOpen
  \bibfield  {author} {\bibinfo {author} {\bibfnamefont {H.}~\bibnamefont {Labuhn}}, \bibinfo {author} {\bibfnamefont {D.}~\bibnamefont {Barredo}}, \bibinfo {author} {\bibfnamefont {S.}~\bibnamefont {Ravets}}, \bibinfo {author} {\bibfnamefont {S.}~\bibnamefont {de~Leseleuc}}, \bibinfo {author} {\bibfnamefont {T.}~\bibnamefont {Macri}}, \bibinfo {author} {\bibfnamefont {T.}~\bibnamefont {Lahaye}}, \ and\ \bibinfo {author} {\bibfnamefont {A.}~\bibnamefont {Browaeys}},\ }\bibfield  {title} {\enquote {\bibinfo {title} {Tunable two-dimensional arrays of single {R}ydberg atoms for realizing quantum ising models},}\ }\href@noop {} {\bibfield  {journal} {\bibinfo  {journal} {Nature}\ }\textbf {\bibinfo {volume} {534}},\ \bibinfo {pages} {667} (\bibinfo {year} {2016})}\BibitemShut {NoStop}%
\bibitem [{\citenamefont {Wang}\ \emph {et~al.}(2020)\citenamefont {Wang}, \citenamefont {Li}, \citenamefont {Feng}, \citenamefont {Song}, \citenamefont {Song}, \citenamefont {Liu}, \citenamefont {Guo}, \citenamefont {Zhang}, \citenamefont {Dong}, \citenamefont {Zheng}, \citenamefont {Wang},\ and\ \citenamefont {Wang}}]{PRL2020}%
  \BibitemOpen
  \bibfield  {author} {\bibinfo {author} {\bibfnamefont {Z.}~\bibnamefont {Wang}}, \bibinfo {author} {\bibfnamefont {H.}~\bibnamefont {Li}}, \bibinfo {author} {\bibfnamefont {W.}~\bibnamefont {Feng}}, \bibinfo {author} {\bibfnamefont {X.}~\bibnamefont {Song}}, \bibinfo {author} {\bibfnamefont {C.}~\bibnamefont {Song}}, \bibinfo {author} {\bibfnamefont {W.}~\bibnamefont {Liu}}, \bibinfo {author} {\bibfnamefont {Q.}~\bibnamefont {Guo}}, \bibinfo {author} {\bibfnamefont {X.}~\bibnamefont {Zhang}}, \bibinfo {author} {\bibfnamefont {H.}~\bibnamefont {Dong}}, \bibinfo {author} {\bibfnamefont {D.}~\bibnamefont {Zheng}}, \bibinfo {author} {\bibfnamefont {H.}~\bibnamefont {Wang}}, \ and\ \bibinfo {author} {\bibfnamefont {D.-W.}\ \bibnamefont {Wang}},\ }\bibfield  {title} {\enquote {\bibinfo {title} {Controllable switching between superradiant and subradiant states in a 10-qubit superconducting circuit},}\ }\href@noop {} {\bibfield  {journal} {\bibinfo  {journal} {Phys. Rev. Lett.}\ }\textbf {\bibinfo {volume} {124}},\
  \bibinfo {pages} {013601} (\bibinfo {year} {2020})}\BibitemShut {NoStop}%
\end{thebibliography}%

\end{document}